# Surface-Plasmon-Polariton (SPP)-Like Acoustic Surface Waves on Elastic Metamaterials

Ke Deng[1,2], Zhaojian He[1,2], Yiqun Ding[1], Heping Zhao[2], and Zhengyou Liu[1,*]

[1]Key Lab of Artificial Micro- and Nano-structures of Ministry of Education and Department of Physics, Wuhan University, Wuhan 430072, China

[2]Department of Physics, Jishou University, Jishou 416000, Hunan, China

We investigate the dispersion properties of the acoustic surface waves on surface of elastic metamaterials. With an analytical approach, we show that unconventional acoustic surface waves, with dispersion behaviors very similar to the electromagnetic surface plasmon polaritons (SPPs) on metal surfaces, can exist on the elastic metamaterials around the frequency at which the elastic Lamè's constants satisfy $\lambda + \mu = 0$. Two typical elastic metamaterials are exemplified to demonstrate such peculiar acoustic surface waves.

The limited properties of conventional matter have been considerably expanded with the development of metamaterials (MMs) during the past few years [1]. The amazing character that (effective) material parameters of MMs can be designed at will has constantly led to conceptual advancements in the fundamental theory of physics, such as the negative medium [2-10], the zero index materials [11-14], the hybrid sold [15], and so on. This in turn has enabled various extraordinary applications such as subwavelengh imaging [16-21], invisibility cloaking [22-25] and enhanced transmitting [26-28], etc.

An important issue involved with MMs is surface waves (SWs) propagating along various interfaces. Electromagnetic (EM) SWs on simple metal-dielectric interface, often called surface plasmon polaritons (SPPs), have been intensively investigated for their ability to control light in subwavelength scale [29]. This ability profits from the very extraordinary characteristics of SPPs that they possess very flat dispersion curves at the characteristic *surface plasmon frequency*. With the generalized parameters from EM metamaterials, now the conventional SPPs have been greatly expanded in terms of frequency band or even polarizing type [30]. Analogously, in the case of acoustic/elastic waves, SWs such as the Stoneley wave are also increasingly receiving renewed interest recently for their potential subwavelength applications [31-33]. However, their applicability is restricted by the limited properties of natural materials. In addition, no conventional acoustic SW possesses the advantages of EM SPPs in terms of spatial localization and high field intensity. From this point of view, investigations of surface waves propagating on acoustic/elastic MMs will be of great significance. Moreover, from a physical point of view, fundamental advancements can also be anticipated with the very unprecedented dispersions relation furnished by MMs.

[*]To whom all correspondence should be addressed, e-mail: zyliu@whu.edu.cn

In this paper, we investigate dispersion properties of SWs propagating on elastic MMs characterized by (effective) mass density $\rho(\omega)$, modulus $E(\omega)$ and $\mu(\omega)$. Here $E \equiv \lambda + 2\mu$, and $\lambda$, $\mu$ are (effective) Lamè's constants. It is a common sense that the longitudinal modulus $E$ must always be larger than the shear modulus $\mu$ for a conventional material, as the Lamè's constants are all positive. This can be different for MMs, of which $E(\omega)$ and/or $\mu(\omega)$ can be negative. Here we consider the type of MMs of which $E(\omega)$ and $\mu(\omega)$ can be equal at some frequency $\omega_0$, $E(\omega_0) = \mu(\omega_0)$, and for the convenience of description hereafter we term the MM as equal modulus solid (EMS). It will be shown that, equipped with these generalized material parameters, EMS bears much richer SW dispersion properties compared with their conventional counterparts. In particular, our investigation reveals the occurrence of flat dispersion curves for SW near the equal modulus frequency $\omega_0$ (i.e., the $k$ goes to infinity when $\omega$ approaches $\omega_0$), accompanied by the opening of a SW gap lying above or below, resembling common SPP dispersion curves [34]. Taking these results as guidelines, we present the designs of two typical types of EMS MMs and demonstrate the unconventional acoustic SWs on these MMs by giving their dispersion relations. In what follows, we will examine two kinds of acoustic SWs: the Rayleigh type (SWs propagating on the surface of a semi-infinite elastic MM in vacuum) and the Stoneley type (SWs propagating on the interface between a semi-infinite elastic MM and a semi-infinite conventional liquid with mass density $\rho_1$ and bulk modulus $E_1$).

**The Rayleigh wave:** The dispersion relation of Rayleigh wave is

$$\left(2\mu k^2 - \rho\omega^2\right)^2 - 4\mu^2 k^2 k_t k_l = 0,  \tag{1}$$

where $\omega$ is the angular frequency, $k$ is the wave number of the SW, and

$$k_t = \sqrt{k^2 - \omega^2 \frac{\rho}{\mu}}, \quad k_l = \sqrt{k^2 - \omega^2 \frac{\rho}{E}}.$$

It is interesting to analyze the behavior of Rayleigh wave around the equal modulus condition ($E = \mu$) or in the limit of $E - \mu \to 0^\pm$. For the purpose, let $E - \mu = \delta$, and our strategy is to expand $E$ around $\mu$ in powers of $\delta$ in Eq. (1), and then to extract the conditions for the opening of gap and the occurrence of divergent k for the Rayleigh wave. The details can be found in the Supplemental Material [35], and here we only summarize the results:

For the case of $\{\rho > 0\}$, the Rayleigh wave gap will lie in the region

$$\{0 < E \leq \mu\},  \tag{2a}$$

or

$$\{E \leq \mu < 0\} \tag{2b}$$

in the parameter diagram, and Rayleigh wave $k \to \infty$ (diverged-k) as

$$\{E \to \mu^+\}. \tag{2c}$$

In contrast, for the case of $\{\rho < 0\}$, the Rayleigh wave gap will lie in

$$\{0 < \mu \leq E\}, \tag{3a}$$

or

$$\{\mu \leq E < 0\}, \tag{3b}$$

and $k \to \infty$ when

$$\{E \to \mu^-\}. \tag{3c}$$

These results can be shown schematically in the parameter diagrams, as Fig. 1(a) and (b).

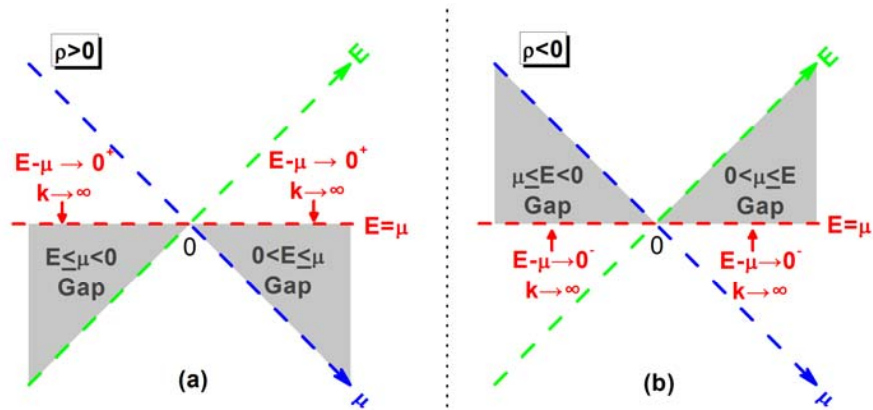

FIG. 1. (color online) Parameter diagrams for the confirmed gap and diverged-k conditions of Rayleigh wave. (a). $\rho > 0$ case, (b). $\rho < 0$ case. Here the gap condition and diverged-k conditions are schematically shown by the shadowed areas and red arrows respectively.

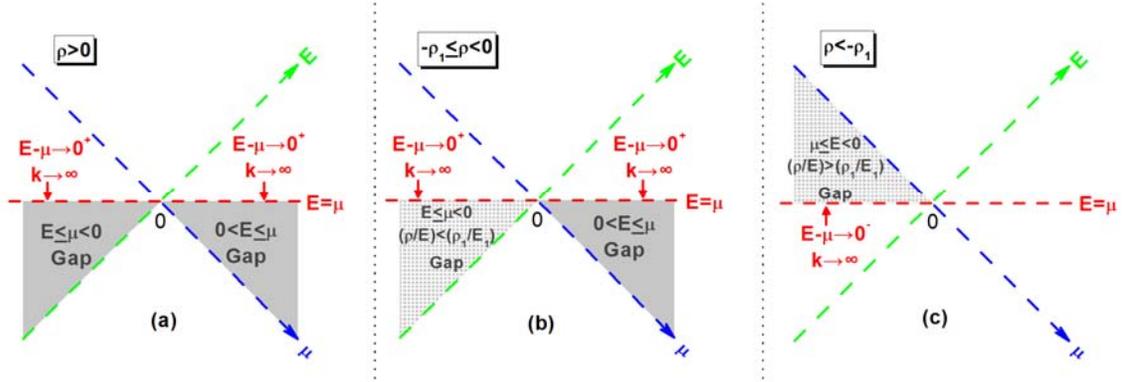

FIG. 2. (color online) Parameter diagrams for the confirmed gap and diverged-k conditions of Stoenley wave. (a). $\rho>0$ case, (b). $-\rho_1 \leq \rho < 0$, (c). $\rho < -\rho_1$ case. Here the shadowed areas represent absolute gaps which are independent of fluid's bulk modulus $E_1$, the patterned shadows represent conditional gaps which are dependent on $E_1$, and the red arrows represent the diverged-k conditions.

**The Stoneley wave:** Dispersion relation of Stoneley SW is

$$k_1 L_R + \rho_1 \rho \omega^4 k_l = 0, \tag{4}$$

with $k_1 = \sqrt{k^2 - \omega^2 \frac{\rho_1}{E_1}}$, $k_l = \sqrt{k^2 - \omega^2 \frac{\rho}{E}}$, and $L_R$ being the left hand of Eq. (1). Again we discuss the behavior of Eq. (4) in the limit of $E - \mu \to 0^{\pm}$ in the parameter diagram. Letting $E - \mu = \delta$, by the same mathematics used in the last section, we can extract the conditions for the opening of gap and the occurrence of diverged-k for the Stoneley wave. Again we leave the details into the Supplemental Material [35], and only summarize the results:

In the case of $\{\rho > 0\}$, the Stoneley wave gap lies in

$$\{0 < E \leq \mu\}, \tag{5a}$$

or

$$\{E \leq \mu < 0\}, \tag{5b}$$

in the parameter diagram, and Stoneley wave $k \to \infty$ as

$$\{E \to \mu^+\}. \tag{5c}$$

In the case of $\{-\rho_1 \leq \rho < 0\}$, the gap lies in

$$\{0 < \mu \leq E\}, \tag{6a}$$

or

$$\left\{E \leq \mu < 0, \text{ subject to } \frac{\rho}{E} < \frac{\rho_1}{E_1}\right\}, \tag{6b}$$

in the parameter diagram, and $k \to \infty$ as

$$\{E \to \mu^+\}. \tag{6c}$$

In the case of $\{\rho < -\rho_1\}$, Stoneley gap lies in

$$\left\{\mu \leq E < 0, \text{ subject to } \frac{\rho}{E} > \frac{\rho_1}{E_1}\right\}, \tag{7a}$$

and $k \to \infty$ as

$$\{E \to \mu^-\}. \tag{7b}$$

A summary of these results for Stoneley wave are schematically shown in the parameter diagrams, Fig. 2(a)-(c).

**Designs of MMs:** As our above studies have shown, there are a variety of conditions for the opening of SW gaps and the occurrence of the SWs with diverged-k, which makes it possible for the appearance of SPP-like flat dispersion curves. To verify this, we have, taking the above results as guidelines, designed two MMs to act as proof-of-concept demonstrations. The first MM is composed of an FCC array of bubble-contained water spheres in epoxy matrix. The filling fraction of the FCC lattice is $26.2\%$ and the ratio of the radii of the air bubble to the water sphere is $2/25$. As has been confirmed [5], this structural unit with monopolar resonance can lead to abnormal bulk modulus in a locally resonant MM. In Fig. 3(a) we give effective parameters of the designed MM calculated by the CPA method [36]. Here, these parameters are normalized by water. Since $\rho$ and $E$ of this MM remains positive throughout the entire frequency range, we should refer to the right part of Fig. 1(a) and Fig. 2(a). The equal modulus frequency for this MM is realized at $\omega_0 = 0.24$. According to Eq. (2) [see Fig. 1(a)] and Eq. (5) [see Fig. 2(a)], gap condition $0 < E < \mu$ is realized in the frequency range marked as $I$, and the SW solution $k$ diverges as $\omega \to \omega_0^-$, both for the Rayleigh wave and for the Stoneley waves. This gives rise to SWs of SPPs-type below the equal modulus frequency $\omega_0$. To verify this, we compute according to Eq. (1) and (4) the SW dispersion relations of Rayleigh wave and of several Stoneley waves on the designed MM, showing the dispersion curves in Fig. 3(b). As expected, there exists a common SW gap in $I$; in addition, when approaching $\omega_0$ from the low frequency direction, all the dispersion curves become very flat and asymptotically reaches infinity, exhibiting behaviors very similar to that of EM SPPs.

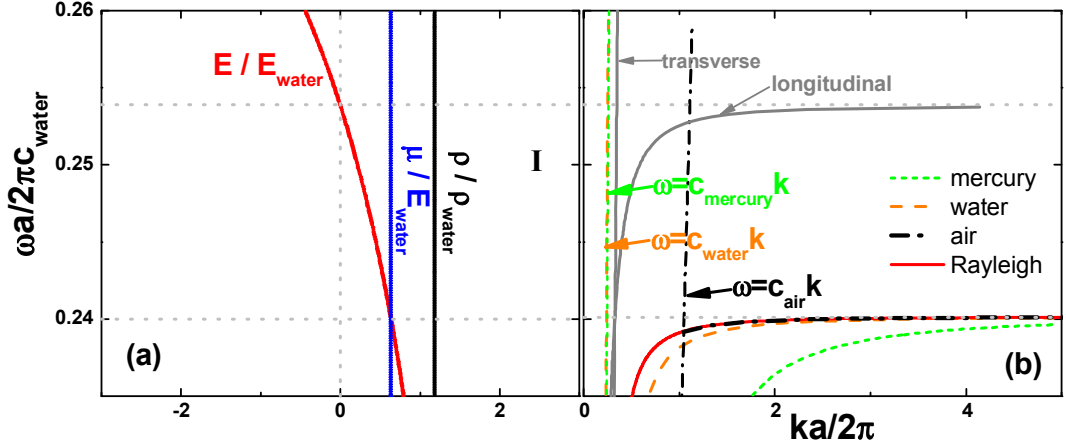

FIG. 3. (color online) (a). Effective parameters of the first designed MM, normalized by the parameters of water. Material parameters used in the calculation are the same as those in refs [5] and [18]. (b). SW dispersion curves for Rayleigh wave (red line) and for Stoneley wave with mercury (short dash), water (dash) and air (dash dot). The dispersion lines for air, water, mercury, longitudinal and transverse waves in MM are also presented.

The second MM is designed by combining an FCC array of bubble-contained water spheres with a relatively shifted FCC array of rubber-coated gold spheres in epoxy matrix to form a zinc blende structure. The filling fraction of the first FCC lattice is $26.2\%$ and the ratio of the radii of the air bubble to the water sphere is $2/25$. The filling fraction of the second FCC lattice is $9.77\%$ and the rubber coating has an inner-to-outer radius ratio of $13/18$. As confirmed previously [5,18], these two structural units with monopolar and dipolar resonances can lead to, respectively, abnormal bulk modulus and abnormal mass density simultaneously in a locally resonant MM. Figure 4(a) gives effective parameters for the second MM calculated by CPA. Here these parameters are also normalized by water. We are mainly interested in the frequency domains marked as $I$ and $I'$ in Fig. 4(a), in which $\rho$ is negative and both $E$ and $\mu$ are positive. Therefore, for this MM we should refer to the right part of Fig. 1(b) and Fig. 2(b). One sees that the equal modulus frequency is realized at $\omega_0 = 0.24$, the condition $0 < E < \mu$ is realized in $I$, and $0 < \mu < E$ is realized in $I'$. For the case of Rayleigh wave, according to Eq. (3) [see Fig. 1(b)], the gap condition is realized in $I'$ and the solution $k$ diverges as $\omega \to \omega_0^+$, this leading to a SPP-like Rayleigh wave above $\omega_0$. To verify this, we compute the Rayleigh dispersion curve for the designed MM as depicted by the red line in Fig. 4(b). One can see that it shows a very flat shape above $\omega_0$, very similar to the EM SPPs curve. For the case of Stoneley wave, the extra requirement of $\frac{\rho}{\rho_1} \geq -1$ should be satisfied to guarantee a gap [see Fig. 2(b) and (c)]. Here we should point out that this MM is elaborately designed such that it satisfies the condition $\left(\frac{\rho}{\rho_{water}}\right)_{\omega_0} = -1$. Therefore, according to Eq. (6) [see Fig 2(b)], $I$ will be a gap and the solution $k$ will diverge as $\omega \to \omega_0^-$ for all cases with $\frac{\rho_1}{\rho_{water}} \geq 1$. This indicates that Stoneley waves on the second EMS will be SPP-type under $\omega_0$ for water and for any fluid which is heavier than water. To verify this, we compute the Stoneley dispersion curves of water and of mercury ($\rho_1 = 13.5 \rho_{water}$) as depicted by the dash and the short dash in Fig. 4(b) respectively. As expected, we obtain dispersion curves with behaviors very similar to that of EM SPPs under the equal modulus frequency. On the other hand, if $\frac{\rho}{\rho_1} < -1$, no Stoneley gap is guaranteed for the second EMS throughout the frequency domain in Fig. 4(a). There exists a finite solution for Eq. (4) at the equal modulus frequency:

$$k = \omega_0 \sqrt{\frac{\rho \rho_1 (\rho E - \rho_1 E_1)}{(\rho^2 - \rho_1^2) E_1 E}}. \tag{8}$$

However, in the limit of $\omega \to \omega_0^+$, the Stoneley wave eq. (4) will reduce to

$$\frac{\rho - \frac{2k^2 \delta}{\omega^2}}{\rho_1} = -\frac{\sqrt{k^2 - \omega^2 \frac{\rho}{E}}}{\sqrt{k^2 - \omega^2 \frac{\rho_1}{E_1}}},$$ with $\delta \to 0^-$. It is easy to see that $k \to \infty$ is always a

solution in this case. This indicates that for those fluids with $\frac{\rho_1}{\rho_{water}} < 1$, Stoneley waves on the second MM will also be SPP-type above $\omega_0$, just similar to the Rayleigh wave. To verify this analyses, we compute the Stoneley dispersion curve for air ($\rho_1 = 0.00129 \rho_{water}$) as depicted by the dash dot in Fig. 4(b) which exhibits SPP-type dispersive behavior above the equal modulus frequency.

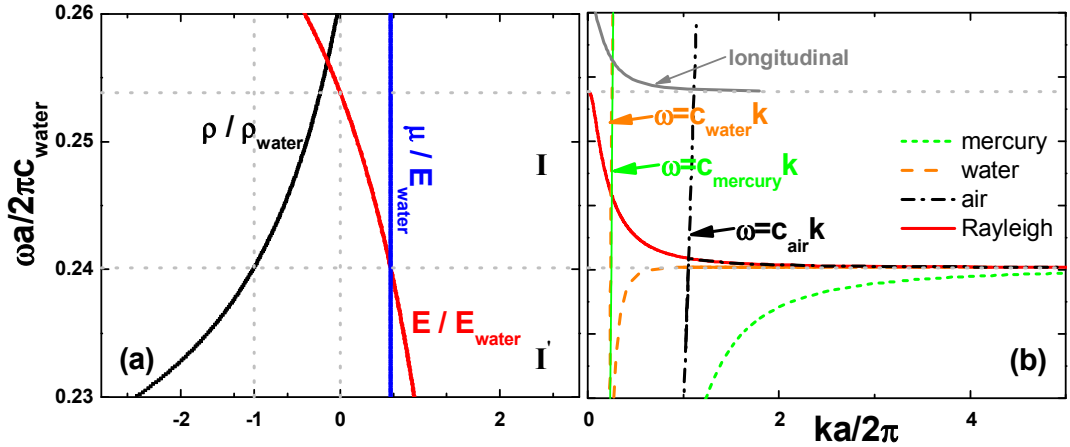

FIG. 4. (color online) (a). Effective parameters of the second designed MM, normalized by the parameters of water. (b). SW dispersion curves for Rayleigh wave (red line) and for Stoneley waves with mercury (short dash), water (dash) and air (dash dot). The dispersion lines for air, water, and mercury and Longitudinal wave in MM are also presented.

**Discussions:** It is easy to find that Eq. (4) reduces to Eq. (1) when mass density $\rho_1$ vanishes. Thus the Stoneley dispersion curve should transit into the Rayleigh one when the mass density of the liquid $\rho_1 \to 0$. This is verified for the designed MMs as one sees in Fig. 3(b) and Fig. 4(b) that the air curves is very close to the Rayleigh curve as we expected.

For the first MM, in a frequency range around $\omega_0$, the parameters of the MM are all positive. This means that there also exist the bulk modes (longitudinal and transverse waves here) in the solid in the frequency range in addition to the SWs, as depicted in Fig. 3(b). For the second MM, however, in the frequency range around $\omega_0$, only the moduli of the MM are positive while the mass density is negative, there are no bulk modes supported in the solid. We see that the Stoneley curves in Fig. 4(b) can be divided into two regimes according to the ratio $\left(\frac{\rho}{\rho_1}\right)_{\omega_0}$ : one with $\left(\frac{\rho}{\rho_1}\right)_{\omega_0} \geq -1$ lying below $\omega_0$ (water and mercury, for example); and the other with $\left(\frac{\rho}{\rho_1}\right)_{\omega_0} < -1$ lying above $\omega_0$ (air, for example, and Rayleigh wave as a limit).

It is well known that the EM (TM polarized) and acoustic equations are exactly equivalent in two dimensions. Ambati *et al* noticed this equivalence and first pointed out the existence of the SPP-like acoustic SW on the interface of two fluids with opposite mass density [17]. They also determined that the negative mass density is a necessary condition for the existence of SWs on such interfaces. On a solid interface, however, SWs can exist without requiring a negative mass density. Here furthermore, we demonstrate that such SWs can also be managed to have their dispersion properties very similar to the EM SPPs. The key is to design MMs with abnormal Lamè's constants satisfying $\lambda + \mu = 0$. As we have shown above, for the designed MMs, dispersion curves are SPP-like both for Rayleigh wave and for Stoneley wave (no matter what liquid adjoined together). These SW dispersion curves are characterized by the same behaviors at frequency $\omega_0$ with $\lambda + \mu = 0$. Compared to SWs on liquid-liquid interfaces, SWs on solid ones take the advantages in applications under various conditions.

In conclusions, we have investigated the dispersion properties of Rayleigh and Stoneley surface waves on the elastic MMs with Lame's constants satisfying $\lambda + \mu = 0$ at certain frequency, which is common for elastic metamaterials designed to exhibit negative modulus. It has been shown that such MM solids support much richer surface wave dispersion relations compared with their conventional counterparts. With an analytical approach, our studies reveal the opening of SW gaps as well as the occurrence of the SW with diverged wave number near the frequency with $\lambda + \mu = 0$. With such dispersion behaviors, the acoustic SWs on such MMs could be managed to mimic the EM SPPs. We think such unconventional acoustic SWs will find promise in subwavelength applications, just as their EM counterpart has done in the rapid expansion of Plasmonics. This work is believed to shed some light on the tailoring of fundamental matter properties and on the designing of novel material applications by metamaterials.

This work is supported by the National Natural Science Foundation of China (Grant Nos. 11374233, 11104113, 11174225, 11264011, and 11304119) and Natural Science Foundation of Education Department of Hunan Province, China (Grant No. 13B091 and 13A077).